# Hexagonal polymorphism induced structural disorder and dielectric anomalies of Ca/Mn modified BaTiO$_3$

P. Maneesha[1], Dilip Sasmal[1], Rakhi Saha[1], Kiran Baraik[2], Soma Banik[2], R. Mittal[3,4], Mayanak K. Gupta[3,4], Abdelkrim Mekki[5,6], Khalil Harrabi[5,7], Somaditya Sen[1]*

[1]*Department of Physics, Indian Institute of Technology Indore, Indore, 453552, India*

[2]*Beamline Development & Application Section, Bhabha Atomic Research Centre, Trombay, Mumbai 400085*

[3]*Solid State Physics Division, Babha Atomic Research Centre, Mumbai 400085, India*

[4]*Homi Bhabha National Institute, Anushaktinagar, Mumbai 400094, India*

[5]*Department of Physics, King Fahd University of Petroleum & Minerals Dhahran, 31261, Saudi Arabia*

[6]*Interdiciplinary Research Center (IRC) for Advanced Material, King Fahd University of Petroleum & Minerals, Dhahran 31261, Saudi Arabia*

[7]*Interdisciplinary Research Center (RC) for Intelligent Secure Systems, King Fahd University of Petroleum & Minerals, Dhahran 31261, Saudi Arabia*

*Corresponding author: sens@iiti.ac.in

## Abstract

This work involves the local structural investigation of the samples using Extended X-ray Absorption Spectra (EXAFS) analysis to investigate structural changes due to the Ca and Mn-modified BaTiO$_3$. TEM investigation of the crystal structure reveals the coexistence of the tetragonal and hexagonal phases of BaTiO$_3$. Band gap modification and Urbach tail variation in UV-DRS measurements reveal a reduction of band gap from UV to Visible range (3.2 eV to 2 eV). This band gap change is correlated with the valence band modification observed in PES measurements due to localised defects in the material, and supported by theoretical Density of States calculations. The electron localisation function calculation is used to analyse the changes in the localised electron density near the dopant atoms. It reveals the local expansion and contraction of the lattice surrounding the dopant atom. All these structural modifications lead to variations in the dielectric properties and diffusive nature of the phase transition. The defect-induced structural modifications, multiple phase coexistence, band gap variations and dielectric properties are explored and correlated.

## Introduction

Crystal structure is an inevitable part of determining multiple properties exhibited by different crystalline materials. BaTiO$_3$, one among the good ferroelectric materials, has been intensively studied for its multiferroic properties by doping it with Transition metal (TM) elements. With the modification, there may or may not be multiferroicity; however, these dopings can play an important role in the manipulation of the disorder-induced electronic, dielectric, and ac conductivity of these materials. Depending on how the disorder influences these physical properties can be effectively used for various multifunctional applications like sensors, actuators, antennas, optoelectronic devices, etc.

The modification of $BaTiO_3$ through A-site and B-site doping is a widely explored method to tune its band gap for various applications, including optoelectronics and photovoltaics [1], [2], [3]. Its wide band gap (~3.2 eV) can hinder its application in fields like ferroelectric photovoltaics, as it limits the absorption of a significant portion of the solar spectrum. Substituting $Sr^{2+}$ for $Ba^{2+}$ in BTO ($Ba_{1-x}Sr_xTiO_3$) can lead to a reduction in the band gap [4]. Experimental and theoretical studies on these ceramics have shown variation in the energy band gap from 3.615 eV to 3.212 eV with Sr substitution, correlating with crystal symmetry transition. Doping Sn at the [5] A-site of BTO can generate impurity bands that alter the direct bandgap to an indirect one and decrease its value, reducing the electron excitation energy. $(Ba_{0.875}Sn_{0.125})TiO_3$ exhibits p-type semiconductor characteristics, which improve the conductivity of BTO. The co-substitution of $La^{3+}$ and $Na^+$ in BTO ceramics, forming $Ba_{(1-x)}(La,Na)_xTiO_3$ (BLNT), introduces structural disorder and A-site vacancies, which leads to the creation of shallow defects and a narrowing of the band gap energy [6]. The replacement of transition metal ions at B-sites in BTO-based solid solutions can lead to a significant decrease in band gaps. Doping BTO with Fe ($BaTi_{0.88}Fe_{0.12}O_{3-\delta}$) results in a smaller band gap compared to pure BTO. Fe doping also creates donor impurity levels in the forbidden band due to the presence of $Fe^{3+}$ and $Fe^{4+}$ forms and positively-charged $O_V$s. With Mn and Nb doping, the band gap of BTO can decrease from 3.2 eV to approximately 2.7 eV for a 7.5% doped sample [7]. Experimental studies have observed a redshift in UV-absorption spectra, indicating a decrease in the optical band gap from 3.13 eV to 2.71 eV as $Mn^{2+}$ doping increases [8]. Apart from the band structure modification, doping can change the $T_c$ and the permittivity and make a diffuse kind of phase transition from ferroelectric to paraelectric. La doping in BTO shifts the Curie point to lower temperatures and controls grain growth efficiently. The dielectric constant of La-doped $BaTiO_3$ ceramics increased with increasing La content [9]. Uniform distribution of $Ca^{2+}$ ions in the BTO matrix makes the transition from the ferroelectric to the paraelectric phase more diffuse and free from relaxational effects up to 10 KHz [10].

Structural studies of Ca and Mn-modified BaTiO$_3$ published elsewhere [11] indicate a pure Tetragonal *P4mm* structure for the x=0 sample and multiple phases of major tetragonal P4mm and minor Hexagonal *P6$_3$/mmc* for x=0.03, 0.06, and majority Hexagonal *P6$_3$/mmc* with minor tetragonal *P4mm* for x=0.09. The multiferroic and magnetoelectric studies revealed that x=0.03 exhibits magnetoelectric coupling at room temperature. The variation of the multiple oxidation states of $Ti^{3+}$, $Ti^{4+}$, $Mn^{3+}$, and $Mn^{4+}$ and the structural changes lead to the magnetoelectric behaviour exhibited by this material.

This study involves the local structural investigation of the samples using Extended X-ray Absorption Spectra (EXAFS) analysis to investigate the local structural changes due to doping at the Mn and Ti sites. TEM investigation of the crystal structure variation with doping. Band gap modification and Urbach tail variation in UV-DRS and changes in the valence band observed in PES measurement due to localised defects in the material, and supported with theoretical Density of States calculation. The electron localisation function calculation is used to analyse the changes in the localised electron density near the dopant atoms. Variations in dielectric properties and changes in phase transitions correspond to these structural modifications, and defect formation is also explored in this work.

**Methodology**

The sample synthesis has been discussed elsewhere [11]. HR-TEM has been done with JEOL JEM F200, King Fahd University of Petroleum & Minerals, Dhahran. The instrument has a Cu-coated carbon grid mesh 400, with a cold field emission gun. The standard normalisation and background subtraction procedures were executed using ATHENA software version 0.9.26 to obtain normalised EXAFS spectra [12]. Valence Band studies have been done with Angle Resolved Photo Electron Spectroscopy (ARPES) Beamline (BL-10) of INDUS-2 Synchrotron Source, Raja Ramanna Centre for Advanced Technology (RRCAT), Indore. Dielectric studies have been done with PSMCOM1735, N4L Impedance Analyser Interface. The Extended X-Ray Absorption Fine Structure (EXAFS) measurement of the K-edges Ti (4966 eV) and Mn (6539 eV) was recorded at the Scanning EXAFS Beamline (BL-09) of the INDUS-2 Synchrotron Source, Raja Ramanna Centre for Advanced Technology (RRCAT), Indore, India. The data was collected when the synchrotron source 2.5 GeV ring was operated at a 120 mA injection current in transmission mode for the Ti K-edge and in fluorescence mode for the Mn K-edge at room temperature

First-principles calculations were performed using the Vienna Ab initio Simulation Package (VASP) within the framework of density functional theory (DFT) [13]. The projector augmented-wave (PAW) method was employed with the Perdew–Burke–Ernzerhof (PBE) exchange-correlation functional [14]. The pseudopotentials used include those for Ca ($3s^2\ 3p^6\ 4s^2$), Ba ($5s^2\ 5p^6\ 5d^{0.01}\ 6s^{1.99}$), Mn ($3d^6\ 4s^1$), Ti ($3d^3\ 4s^1$), and O ($2s^2\ 2p^4$), ensuring accurate treatment of core and valence electrons. A high plane-wave energy cutoff of 700 eV was used to ensure convergence of total energy and electronic properties. The electronic self-consistency loop was converged with a stringent criterion $1.0E^{-7}$ eV, while ionic relaxation was controlled using $1.0E^{-3}$ eV/Å, indicating force-based convergence. Structural optimisation was carried out using the conjugate gradient algorithm with full relaxation of atomic positions and cell parameters. Spin-polarised calculations were performed for all the cases. Brillouin zone sampling was done using a Monkhorst-Pack grid of 4×4×4 [15]. Post-relaxation, the band structure was calculated along high-symmetry paths to analyse electronic dispersion. The density of states (DOS) was computed using a dense k-grid of 10x10x10 to examine the distribution of electronic states across energy levels, and the electronic localisation function (ELF) was evaluated to visualise bonding characteristics.

## Results and Discussion

**Transmission Electron Microscopy**

High Resolution TEM image of all samples is shown in Figure 1. The BTO sample shows a long-range tetragonal lattice in (111) directions with a d spacing of 2.31 Å. The selected area electron diffraction pattern with bright spots in a ring pattern indicates the reflections from different zone axes of the tetragonal lattice structure. In the x=0.03 sample, there are two kinds of lattice structure corresponding to the tetragonal and hexagonal lattices. The (103) planes of a hexagonal lattice with d spacing 3.4 Å are parallel with the (102) planes of a tetragonal lattice with d spacing 1.8 Å. The boundary between the two lattices indicates a deviation from the lattice planes of the tetragonal and hexagonal lattices. x=0.06 sample shows the coexistence of parallel tetragonal (002) (d spacing 2.01 Å) planes and hexagonal (006) (d spacing 2.01 Å) planes. x=0.09 shows the tetragonal (103) (d spacing 3.25 Å), (003) (d spacing 1.32 Å) and hexagonal (104) planes.

SAED pattern of the x=0.09 hexagonal region shows blurred spots with ring patterns for different lattice planes of the hexagonal lattice. The existence of both lattice structures in doped samples indicates that Ca and Mn doping stabilise the hexagonal structure in tetragonal BTO. The TEM image shows how the hexagonal and tetragonal lattices are interlinked with each other, and the enhancement of hexagonal lattice formation with doping.

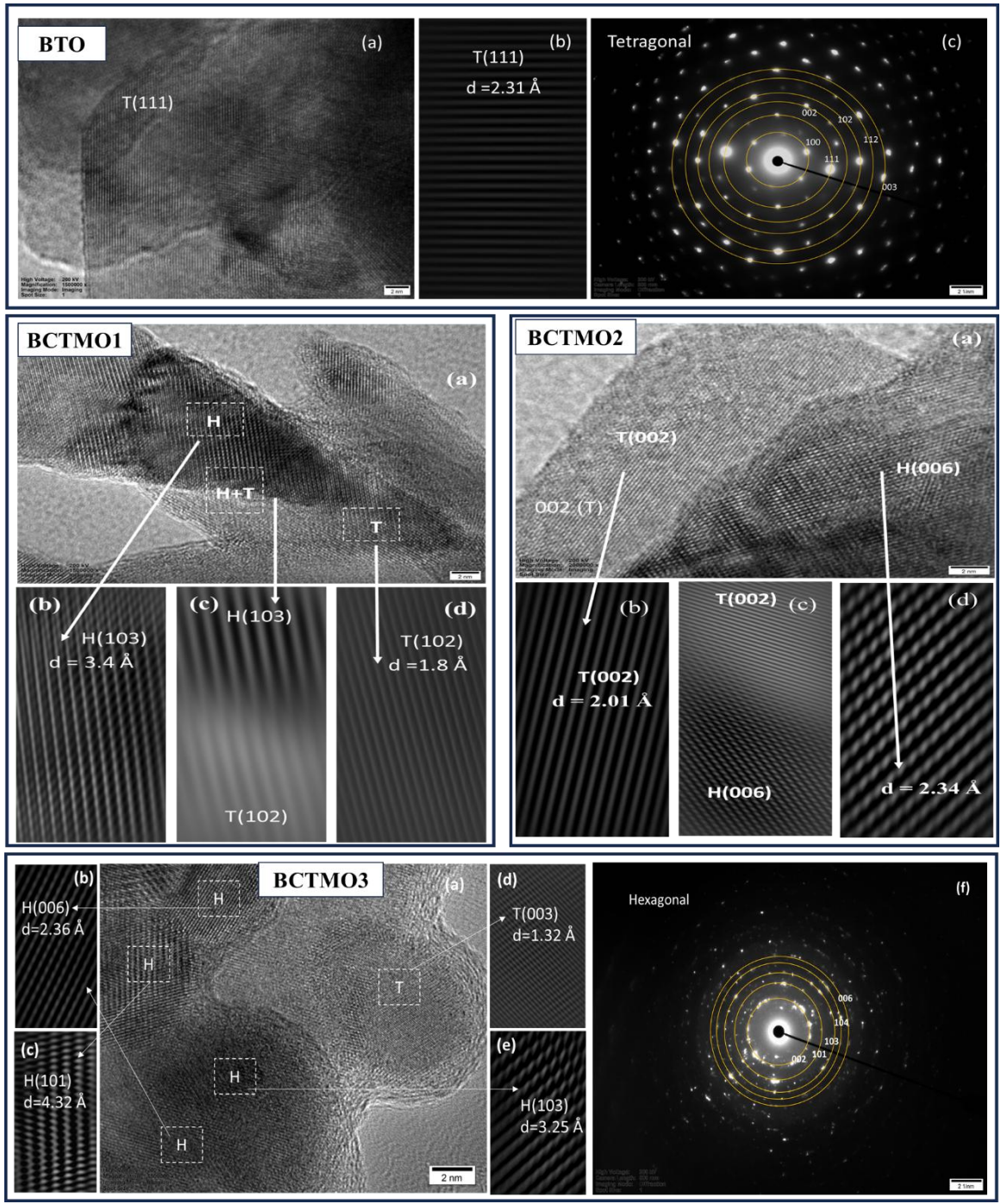

*Figure 1:* x=0 →(a) TEM image of tetragonal lattice, (b) back ground subtracted lattice plane image of (a), SAED pattern of (a), x=0.03 → (a) TEM image of hexagonal and tetragonal lattice, (b),(c),(d) back ground subtracted lattice plane image of selected region. x=0.06 → (a) TEM image of hexagonal and tetragonal lattice, (b),(c),(d) background-subtracted lattice plane image of selected region. x=0.09→ (a) TEM image of hexagonal and tetragonal lattice, (b),(c),(d),(e) background-subtracted lattice plane image of selected region, (f) SAED pattern corresponding to the hexagonal region.

**X-Ray Absorption Spectroscopy Analysis**

EXAFS spectra provide the scattering paths that include local structural changes surrounding the atom. Ti K-edge EXAFS for the samples is shown in Figure 2(a). The radial distance of the shell configuration, which is assignable to scattering paths of Ti-O, Ti−Ba, and Ti−O−Ti [16]. All three peaks shift towards a lower position with doping, corresponding to the decrease in the Ti-O, Ti-Ba, and Ti-Ti distances. An increase in the intensity of the EXAFS oscillation corresponding to the Ti-O bond with doping indicates the oxygen environment surrounding the Ti is less distorted. This indicates the centrosymmetric nature of Ti in the doped samples. Intensity of the Ti-Ba path shows a random behaviour with doping, while the Ti-O-Ti scattering path shows an increase in intensity of the peak up to 0.06 and then a decrease for x=0.09. For all doped samples, the intensity of the Ti-O-Ti path is more than the x=0 sample, indicating a less distorted Ti-O-Ti path with doping.

Mn EXAFS spectra [Figure 2(b)] show the main feature at 1.35 Å, related to the first oxygen coordination shell (Mn–O) for all samples. The position of this peak slightly shifts to higher values with increasing Mn content. The enlargement of the Mn–O bond length with increasing Mn content is indicative of decreasing larger ionic radius $Mn^{3+}$ proportions over smaller ionic radius $Mn^{4+}$. The multipeak at ~3.2 Å corresponds to the second (Ba) and third near neighbours (Ti, Mn, and Nb), including multiple scattering contributions from Mn−Ti−O paths [17]. This peak corresponds to a weighted sum of contributions from Mn−Ti, Mn−Mn, and Mn−Ba. This peak appears shifted towards lower values, corresponding to the decrease in the Mn-Ba/Ti/Mn bond with doping. The Mn−O peak amplitude tends to increase with doping, indicating that the oxygen environment around the manganese is becoming less distorted, indicative of the transformation to more centrosymmetric, less tetragonal behaviour with doping [17]. However, the multipeak the intensity decreases gradually with doping, indicating the higher disorder of the second shell with doping. The contraction of the first shell and the expansion of the second shell, opposite changes in the variations in the intensity of the scattering paths of the first and second shell, reveal the modifications in the coordination sphere of Mn and thus the crystal structure changes with doping [18]. The third peak observed ~3.7 Å for the x=0.03 sample appears much smoother in other samples, which suggests the distribution of interatomic distances for that shell is possibly more disordered with doping and not significantly observed in the higher-doped sample.

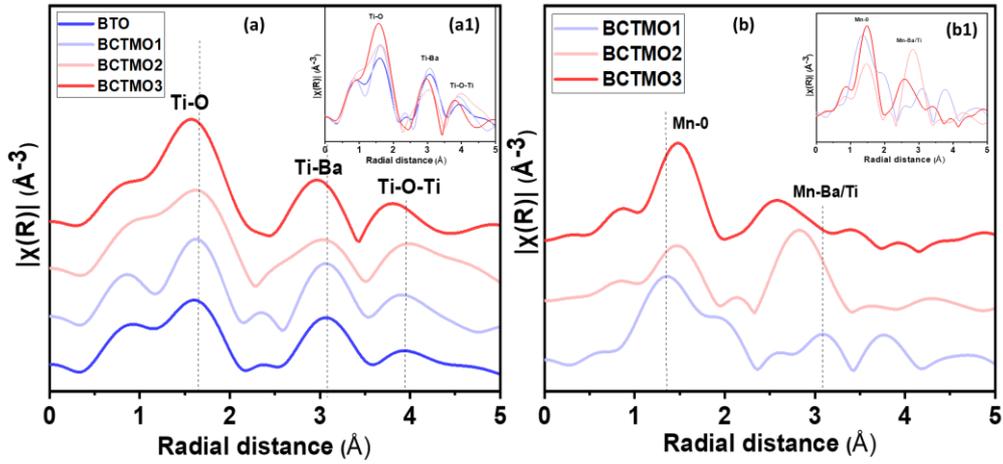

*Figure 2: Fourier transformed EXAFS spectra of (a)Ti edge, (b) Mn edge*

**UV-Visible DRS analysis**

The emergence of the hexagonal phase can induce band gap variations due to the defects and disorder created in the material. The defect states created can effectively modify the band gap energy ($E_g$). The band tails, which can be a material property of an imperfect semiconductor or can be introduced into the material by doping. The exponential Urbach tail at the absorption edge results from transitions between these band tails below the band edges and changes an abrupt absorption to a bent absorption at the band gap energy. Hence, the analysis of the Urbach energy ($E_U$) indicates the disorder in the material.

$E_g$ was measured using the UV-visible absorption spectrum and is estimated by fitting the absorption coefficient in Tauc's relation given by $(\alpha h\nu)^n = C(h\nu - E_g)$ [26], where C is the proportionality constant, h is the Planck constant, $E_g$ is the optical bandgap, and n is 2 for the direct bandgap [27]. Tauc's plot of all the samples is given in Fig.3(a). The extrapolation of the straight portion of the curve to the energy axis will give the bandgap. The structural disorder and the defects in the lattice are associated with the $E_U$ [28]. The spectral dependence of the absorption coefficient α in the spectral region corresponding to transitions involving the tails of the electronic density of states is described by the Urbach equation,

$$\alpha = \alpha_0 exp\left(\frac{h\nu - E_g}{E_U}\right),$$

where $\alpha_0$ is a constant, $E_U$ is the energy that reflects structural disorder and defects of a semiconductor.

The $E_g$ and $E_U$ are calculated from the $(\alpha h\nu)^n$ v/s $h\nu$ plots. In polycrystalline metal oxides, which have disorder due to doping, the Urbach tail influence on the band gap is very high it can shift the band gap by a few hundred meV. Hence, $(\alpha h\nu)^n$ plots can be used to determine the bandgap of this material only when the influence of disorder on the absorption edge is not too strong. For this purpose, Viezbicke et al. [19] have developed the NEAR factor (Near-Edge Absorptivity Ratio) to describe the influence of band tails on the absorption edge. The NEAR factor for a direct allowed transition is determined using the following formula.

$$\text{NEAR} = \left\{ \frac{(\alpha h)^2|_{h\nu=E_g}}{(\alpha h)^2|_{h\nu=1.02E_g}} \right\} = \frac{\alpha(E_g)}{\alpha(1.02E_g)}$$

NEAR factor is intended to determine how steep the absorption edge is near the bandgap. To do so, the absorption at the energy of the bandgap is compared with the absorption at a photon energy enhanced by 2%, i.e., 1.02 $E_g$. When the Urbach tail is significant, NEAR approaches 1, while when the absorption edge is little affected by the Urbach tail, NEAR is small, ideally approaching 0. The $(\alpha h\nu)^n$ plot should only be used if a NEAR factor < 0.5 can be achieved.

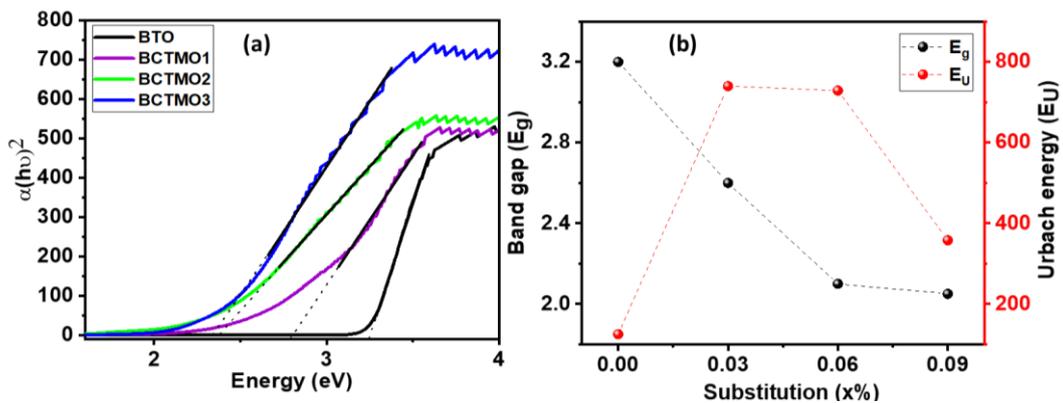

**Figure 3:** *(a) UV-DRS tauc plot of all samples, (b) Variation of $E_g$ and $E_U$ with substitution.*

In all the samples, the NEAR factor is calculated to be <0.5, and hence the influence of Urbach tail on the Band gap is negligible. The Tauc plot of all samples for direct band gap (n=2) is shown in Figure 3(a). The calculated $E_g$ and $E_U$ are plotted in Figure 3(b). $E_g$ is showing a gradual decrement from 3.2 eV to 2.05 eV as the doping increases from x=0 to x=0.09. $E_U$ shows an increment from 125 eV to 740 eV from x=0 to x=0.03, then decreases to 358 for x=0.09.

| Table 2.1: Obtained values of $E_g$ and $E_U$ from UV-DRS |||
|---|---|---|
| Sample | Band gap (eV) | Urbach energy (meV) |
| BTO | 3.2 | 125 |
| BCTMO1 | 2.6 | 740 |
| BCTMO2 | 2.1 | 729 |
| BCTMO3 | 2.0 | 358 |

A sudden decrease in the $E_U$ from x=0.06 to x=0.09 is due to the drastic structural change from a minor hexagonal *P6₃/mmc* phase with major tetragonal *P4mm* in x=0.06 to a minor hexagonal *P6₃/mmc* phase percentage with major tetragonal *P4mm* phase percentage in x=0.09. $E_U$ increases as the disorder in the tetragonal phase increases, and then for x=0.09, there is a complete structural change taking place in which the hexagonal *P6₃/mmc* phase is the major phase and the tetragonal phase is the minor phase. The disorder in the tetragonal phase attains its maximum at x=0.03, and then the crystal structure tries to minimise the strain due to doping and defect formation by triggering the hexagonal phase, which results in a decrease in $E_U$ value. The drastic change observed in the $E_g$ value corresponds to the defect states formed in the band gap due to doping.

**X-Ray Photo Emission Spectroscopy**

The electronic property was investigated using photoemission spectroscopy (PES). The valence band (VB) spectra for all compositions are shown in Figure 4(a). Analysing valence electron distribution provides essential information regarding the physical and chemical properties of the solid. The hybridisation of 2p electrons of oxygen with 3d electrons of Ti contributes to the primary share of the density of state (DOS) of the valence band (VB). The measured VB spectra are an integrated DOS throughout the entire Brillouin zone. For the interpretation of the partial DOS, it should be taken into account that the Ti state in XPS VB spectra demonstrates an overestimated contribution of the occupied 3d state due to a higher cross-section of Ti than for oxygen [20]. The valence band is deconvoluted into 3 Gaussian peaks of three electronic states: one pure O-2p orbital and two O-2p and Ti-3d hybridised states are called as regions A, B and C, respectively, for BTO and for doped samples, two additional ingap states (IGS) named as D and E are also observed [Fig 4(b)] [21].

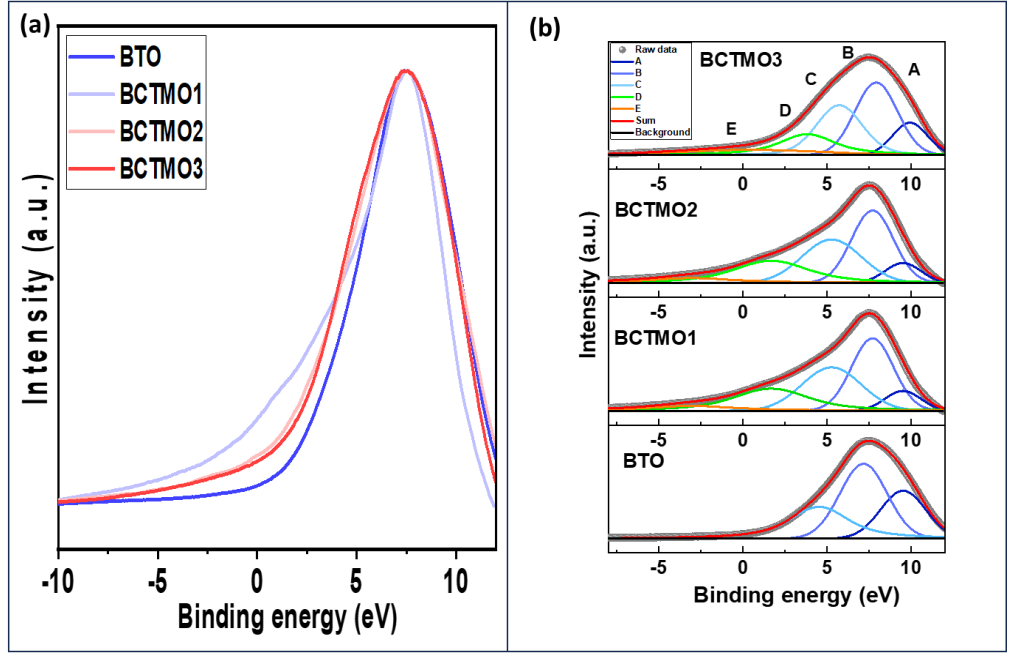

*Figure 4:* *(a)VB spectra of different samples, (b) Deconvoluted VB spectra*

The VB spectra reveal a growing intensity in valence states from 2 eV extending to about 11 eV in binding energy (BE). The onset of the valence band maximum (VBM), defined from the Fermi level ($E_F$) at 0 eV, also sees a clear shift towards lower BE as a function of doping. We interpret this shift in the valence band edge due to admixed dopant *d* states that hybridise with O 2*p,* which push the VBM towards the conduction band and reduce the bandgap [22].

For the doped samples, VB tail rising earlier than x=0 indicates the enhanced IGS in the sample. The IGS may arise due to the vacancies that form the trap centers. Such states are most commonly found to originate from oxygen defects [23], [24] and $Ti^{3+}$ ions located in the vicinity of oxygen defects [25]. IGS are represented by the peaks D and E in the doped samples. Peak E corresponds to the collective effect of the $Ti^{3+}$ defect state and deep trap due to oxygen defects, while peak D corresponds to the shallow trap due to oxygen defects [26]. Variations of the normalised area of the IGS, D and E are plotted in Figure 5. It is evident that the overall area of the peak profile of the IGS peak D is maximum for the BCTMO1 sample, then decreases with doping, whereas peak E increases from BCTMO1 to BCTMO3.

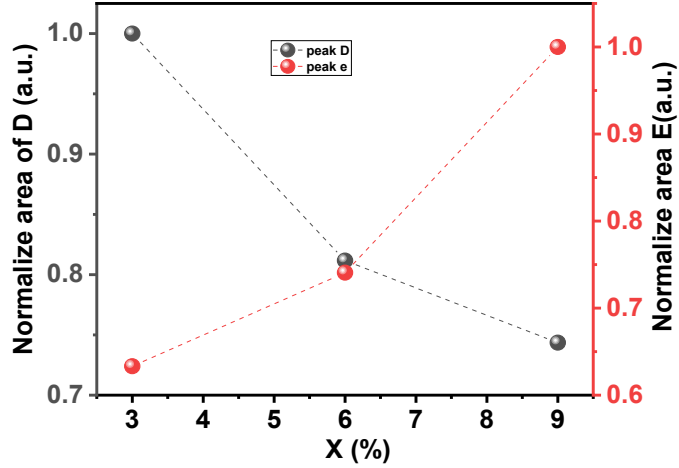

*Figure 5: Variation of the normalised area of IGS of peaks D and E with substitution.*

Hence, the concentration of $O_v$ is maximum in the BCTMO1 sample and thereafter decreases as the structure goes towards a more hexagonal nature. And the defect states $Ti^{3+}$ slowly increases from BCTMO1 to BCTMO3 [Fig. 5]. The valence band tail crosses $E_F$ due to doping-induced localised states. The origin of $O_v$ and $Ti^{3+}$ in BTO triggers the hexagonal phase of $BaTiO_3$, as the hexagonal phase percentages increase to minimise these defect states. Similar results were obtained in the $E_U$ calculation from the UV-DRS spectrum. Urbach energy increases from BTO to BCTMO1, then decreases afterwards. PES and UV-DRS measurement both indicates the defect states within the bandgap that lead to Urbach tails in the valence band in PES and absorption spectrum in DRS.

**Density of States Calculation**

To compare the experimentally obtained band gap modification, theoretical DOS calculations have been performed for x=0 and x=0.125 for tetragonal and hexagonal $BaTiO_3$, and are shown in Figure 6. The valence band of the $BaTiO_3$ is mainly formed by the Ti 3d orbitals, and the conduction band is formed by the O 2p orbital. Bandgap of the tetragonal x=0 sample is obtained as 1.4 eV, Fig. 6(a). DOS calculations of doping of Ca and Mn for x=0.125 in the tetragonal phase are shown in Fig. 6(b). The band gap has reduced to 1.123 eV with doping in the tetragonal phase without any defect states within the band gap. Hexagonal $BaTiO_3$, x=0 sample shows band gap value of 1.053 eV [Fig 6(c)], and doping of Ca and Mn for x=0.125 hexagonal $BaTiO_3$ [Fig 6(d)] becomes a nearly metallic state (band gap ~0eV) with a lot of defect states of Mn and O. In the experimental study, the sample contains a mixture of these hexagonal and tetragonal $BaTiO_3$, with doping possible in both of these phases. Hence, doping in the tetragonal phase leads to the shifting of VB and CB and results in the reduction of the effective bandgap. However, doping in the hexagonal phase can generate Mn and O defect states such as $Mn^{3+}$, $Mn^{4+}$, $O_v$, etc, and effectively reduces the band gap of the material. Both these effects are simultaneously occurring in all BCTMO1, BCTMO2, and BCTMO3 samples, and the cumulative effect of shifting of VB and CB together with the defect states collectively reduces the effective band gap.

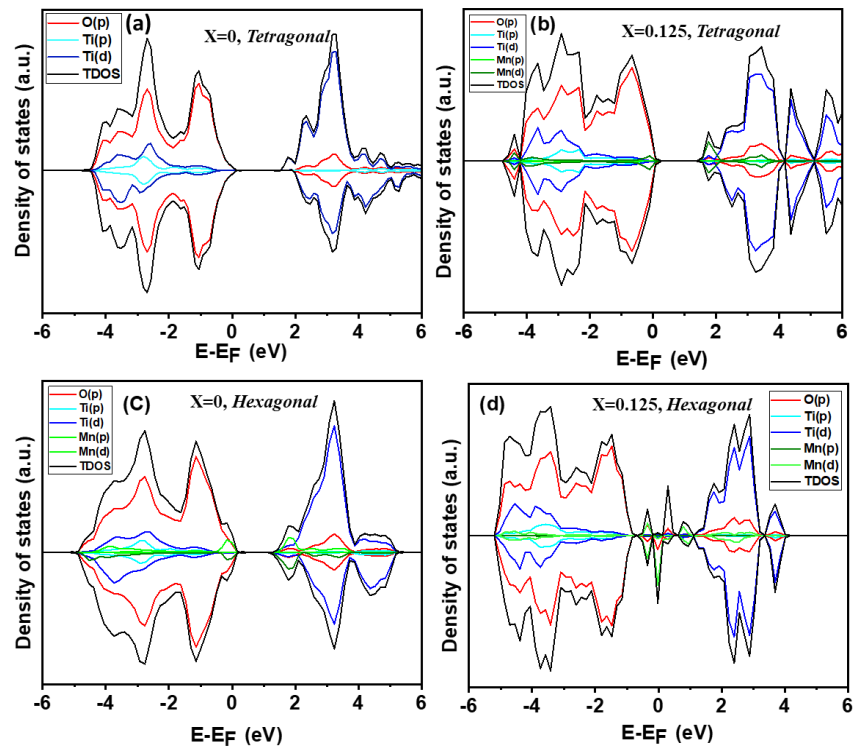

*Figure 6:* (a) DOS plot of x=0 tetragonal, (b) DOS plot of x=0.125 Ca and Mn doped tetragonal, (c) DOS plot of x=0 hexagonal BTO, (a) DOS plot of x=0.125 Ca and Mn doped hexagonal BTO

As observed in PES, Urbach tails due to IGS are forming due to the doping of Mn and Ca in hexagonal $BaTiO_3$, as observed in Fig. 6(d). The defect states are formed due to Mn(d) and O(p) states near the VBM. These defect formations are in correlation with the PES and UV-DRS measurements.

**Electron Localisation Function Calculation**

The local electronic cloud modification created in the tetragonal lattice due to doping and hexagonal structure polymorphism has been studied with Electron Localization Function (ELF) calculation. Fig. 7(a1) shows the ELF plots of the (001) plane of the Ca-doped Ba-O lattice, and Fig. 7(a2) shows the Mn-doped Ti-O lattice in tetragonal $BaTiO_3$ for x=0.125. It has been observed that there is a lattice contraction surrounding the dopant Ca atom. Lattice expansion is observed surrounding to dopant Mn atom. The lattice expansion created due to Mn substitution is not isotropic. Replacement of dimer Ti atoms by Mn atoms in the (-101) plane hexagonal lattice is shown in Figure 7(b1), and that of the (001) plane is shown in Figure 7(b2) for x=0.125. The replacement of centrosymmetric octahedral Ti atoms in the hexagonal lattice by Mn atoms is shown in the (-101) plane in [fig.7 (c1)] and (001) plane [Fig. 7(c2)].

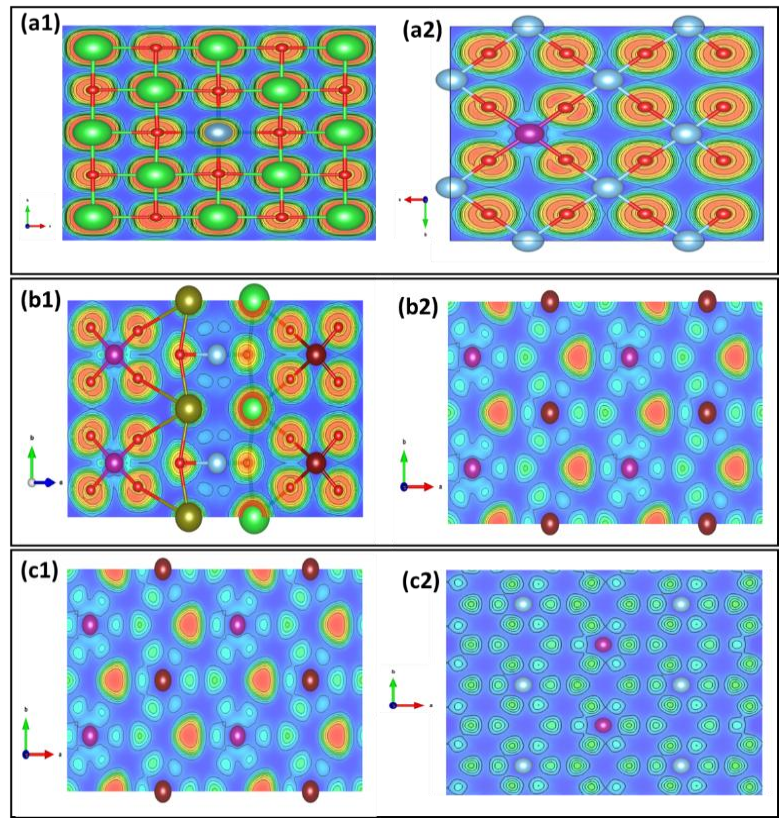

***Figure 7:*** *(001) plane of tetragonal BaTiO$_3$ for x=0.125 (a1) Ca doped Ba-O lattice and (a2) Mn doped Ti-O lattice, Replacement of dimer Ti atoms by Mn atoms in (b1) (-101) plane hexagonal lattice and (b2) in (001) plane is shown in for x=0.125, The replacement of centrosymmetric octahedral Ti atoms in the hexagonal lattice by Mn atom (c1) in the (-101) plane and (c2)(001) plane for x=0.125.*

It has been observed that Mn substitution shows lattice expansion for both octahedral (Ti1) and dimer (Ti2) Ti ions replacement in the hexagonal phase, as observed in tetragonal lattice Mn substitution. This observation is in correlation with the EXAFS analysis of the Mn K edge. Ca substitution of Ba in both lattices shows lattice contraction surrounding the dopant atoms. These lattice modifications are localised changes that occur surrounding the dopant atoms.

The structural modifications observed in EXAFS data and the defect states with doping observed in DRS, PES, DOS, and ELF results indicate that the defects induced a hexagonal lattice in tetragonal BaTiO₃. These modifications can influence the dielectric properties exhibited by these materials.

**FE-SEM Analysis**

FESEM images of optimally sintered pellets show [Fig. 8] a dense micrograin for all samples. The sintering of all the samples at 1330 °C for 4 h forms a crystalline nature with dense matter formation. All the sintered pellets have having relative density of more than 90%. Since the sintering temperature decreases with doping, the doped sample sintered more compared to the pure sample. During the sintering process, crystalline dense grains are formed by the site-to-site diffusion of ions. Vacancies in the lattice play a major role in materialising such an ion diffusion process. The O$_V$ available in the lattice is are important factor in the grain-growth mechanism. Pure BTO shows a cuboid morphology of grains, while BCTMO samples show two morphologies of cuboid one and needle type morphology. This lossy needle-like morphology corresponds to the 110-axial growth of the hexagonal structure along with the tetragonal BTO cuboid grains. The fraction of non-cuboid grains increases with doping, indicating the phase transition of BTO from tetragonal to hexagonal structure.

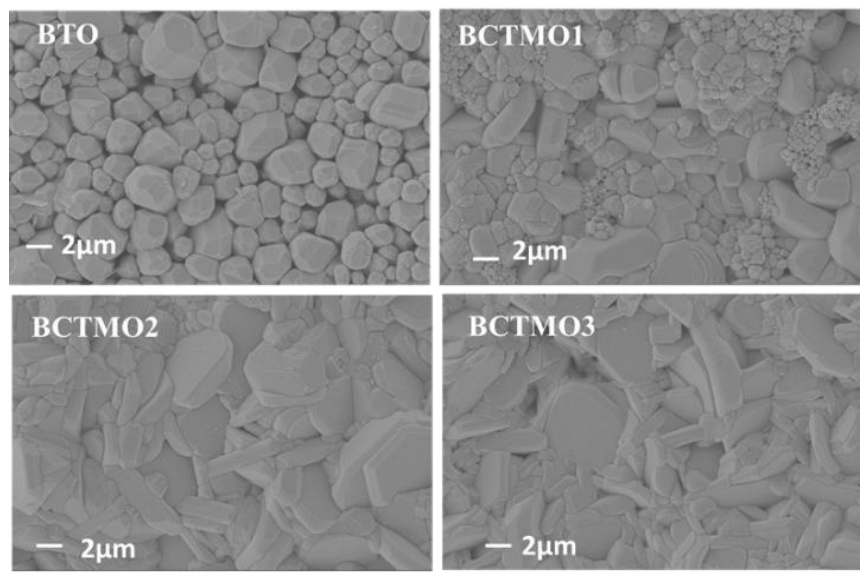

*Figure 8:* *FESEM micrographs of all samples.*

**Dielectric Properties**

Room temperature frequency-dependent dielectric constant (ε$_r$) [Fig. 9(a)] and loss factor (tanδ) [Fig. 9] was studied in detail. There is a decrease in the permittivity value with frequency has been observed, which can be related to the space charge polarisation. The dispersion of permittivity at lower frequencies is more for BCTMO3, indicating an increase in interfacial polarisation that arises from the separation of charge carriers accumulated at the grain boundary and electrode sample interface. From the structural studies, it has been observed that with substitution, the non-centrosymmetric nature of Ti is reduced, and also the paraelectric hexagonal phase percentage increases. The cumulative effect of these factors reduces the effective polarisation within the samples that resulting in the reduction of the permittivity with doping.

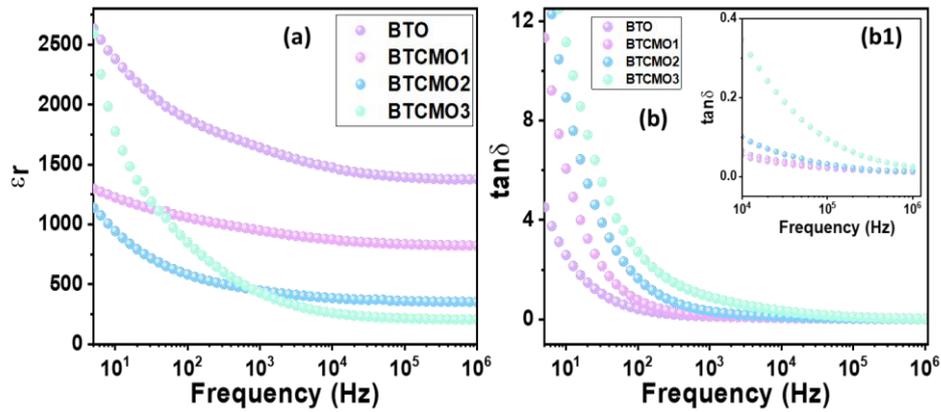

*Figure 9:* *(a)RT permittivity variation with frequency, (b) RT tanδ variation with frequency.*

Conduction losses in the materials are analysed by the measurement of tanδ [Fig.9b], which shows the increment with doping. At lower frequencies, the value of tanδ will be high, and with frequencies, it decreases due to the space charge and interfacial contributions. Inset of figure 9b [Fig.9b1] indicates the variation of tanδ at higher frequencies, which is almost the same for all samples except BCTMO3, which shows the higher dielectric loss even at higher frequency. From the FESEM morphology, the lossy rod-like features increase with an increase in doping percentage, which is also obtained from the dielectric loss measurement. In addition to this, the increase in the presence of multivalent cations, as in $Mn^{3+}/Mn^{4+}$ and $Ti^{3+}/Ti^{4+}$, enables more charge carriers to be freely available and leads to an increase in conductivity.

Variation of permittivity within the temperature range 50 °C to 200 °C for a frequency sweep of 10Hz to 1MHz is shown in Figure 10. All samples show a ferroelectric to paraelectric phase transition indicative of the presence of a non-centrosymmetric ferroelectric *P4mm* phase in these materials. The curie temperature is ~116 °C for BTO while ~104 °C for BCTMO1, ~99 °C for BCTMO2 and ~101 °C for BCTMO3. A decrease in Curie temperature is indicative of the decrease in the non-centrosymmetric behaviour as doping increases. The transition temperature value has no variation as a function of frequency, which confirms that these compositions are not relaxor-type and exhibit classical ferroelectric-type.

A sharp phase transition for BTO is modified into a broader diffuse type phase transition in doped samples. The diffuseness in the Curie temperature is due to the formation of nanopolar regions due to the Ca and Mn doping in BTO.

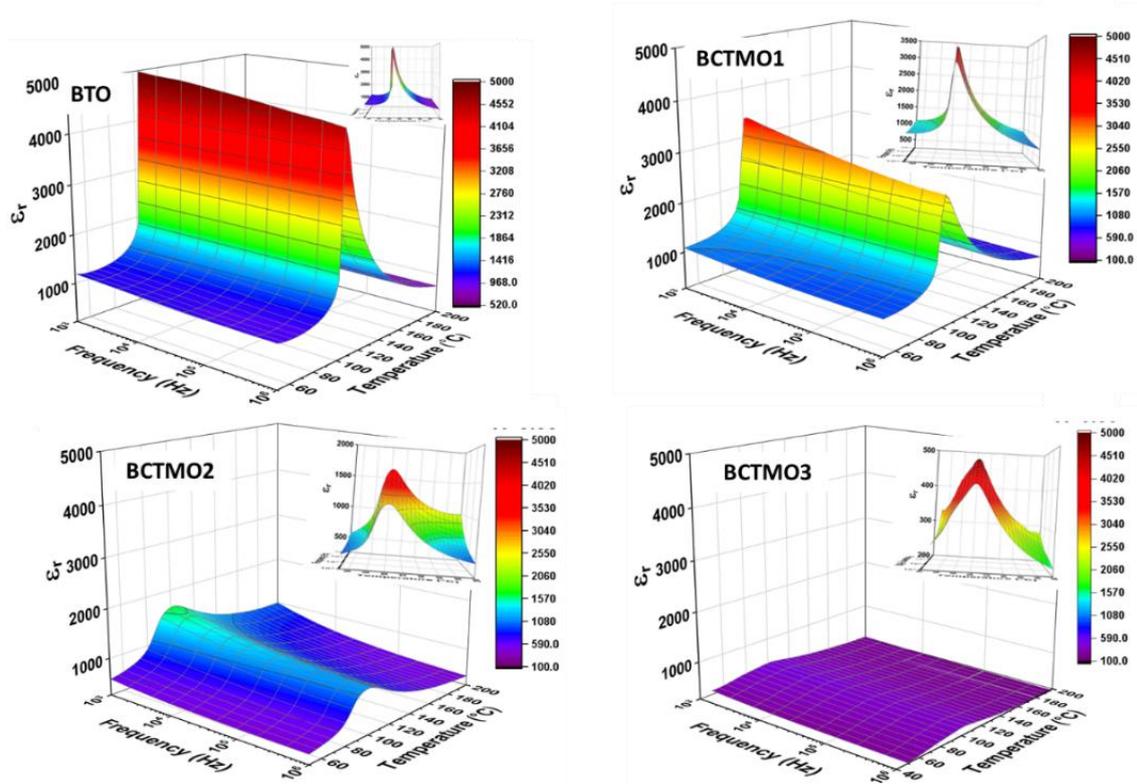

*Figure 10: Temperature-dependent permittivity variation showing ferroelectric to paraelectric phase transition.*

The homogeneous nature of BTO leads to a single $T_c$ for the ferroelectric to paraelectric phase transition. However, doping of Ca and Mn leads to variations in the polarisation in and around the doping sites, which creates polar regions with different polarisations and requires a temperature of phase transition with a distribution of $T_c \pm \delta$, which leads to the diffuse phase transition in the doped samples [Figure 11]. Moreover, the decrease in the non-centrosymmetry with doping leads to a decrease in the polarisation and hence requires a much lower temperature for the phase transition to occur. This implies the reduction of the $T_c$ value with doping.

The dielectric permittivity, $\epsilon'$ of a normal ferroelectric above the Curie temperature follows the Curie–Weiss law described by [27]:

$$\epsilon' = \frac{C}{(T - T_m)} \ for \ T > T_C$$

where $T_m$ is the temperature at which ε has maximum value, $\varepsilon_m$, C is the Curie–Weiss constant, and $T_C$ is the Curie temperature of the ferroelectric to paraelectric phase transition. For ferroelectrics, this expression is no more than a mean-field approximation applied to the fluctuating local electric fields in the crystal structure.

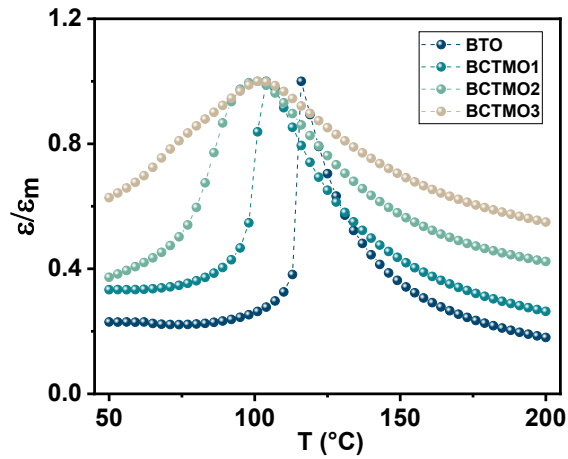

*Figure 11:* *Comparison of the phase transition from a sharp nature to a diffuse nature with doping.*

Figure 12 shows the plot of inverse dielectric constant vs temperature for all samples. All the plots clearly show a deviation from the Curie-Weiss law near the phase transition temperature. For pure $BaTiO_3$, this deviation is taking place abruptly at the $T_c$, while for the doped samples, deviation from the Curie-Weiss law starts a few temperatures above $T_c$ and shows a gradual diffusive nature near the $T_c$.

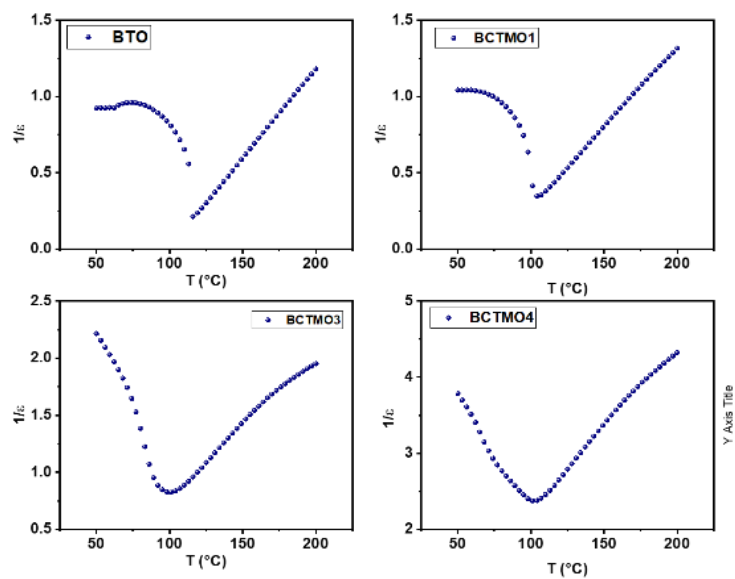

*Figure 12:* *$1/\varepsilon$ v/s T plot of all samples.*

To better describe the dispersion degree of diffuse ferroelectric phase transition, the Curie–Weiss law has been revised and can be described by a modified Curie-Weiss law [27]:

$$\frac{1}{\varepsilon} - \frac{1}{\varepsilon_m} = \frac{(T - T_m)^\gamma}{C}$$

where $\varepsilon_m$ is the dielectric constant (maxima) at $T_m$, $\gamma$ ($1 \leq \gamma \leq 2$) is the degree of diffuseness, and C is the Curie-Weiss constant. For a purely ferroelectric material, $\gamma = 1$, $1 < \gamma < 2$ is a diffuse transition, and for a relaxor ferroelectric, $\gamma = 2$. Figure 13 shows a linear fit of ln $(1/\varepsilon - 1/\varepsilon_m)$ vs ln$(T-T_m)$ plots for different frequencies. The slope of the plots gives the value of $\gamma$.

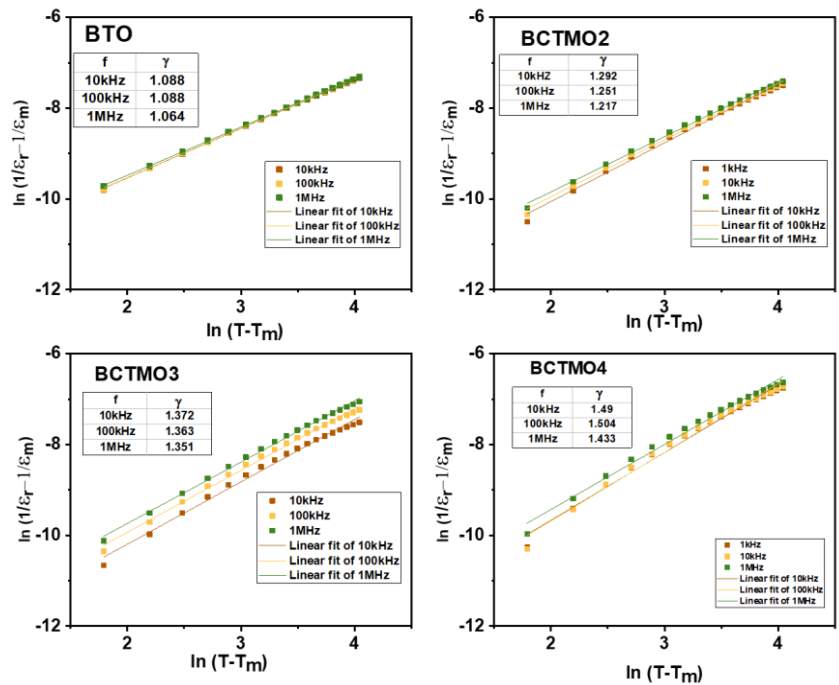

*Figure 13: ln $(1/\varepsilon - 1/\varepsilon_m)$ vs ln$(T-T_m)$ plots for all samples*

All samples have a $\gamma$ value in between 1 and 2 ($1 < \gamma < 2$), and there is no deviation of $\varepsilon_m$ with frequencies, eliminating the possibility of a relaxor nature in these samples. The diffusive behaviour shown by these samples occurs because of the chemical disorder created due to the doping of Ca and Mn in $BaTiO_3$. Due to the random distribution of atoms, localised structural phase transition gives rise to the variation in phase transitions over a broad range of temperatures. The value of $\gamma$ increases with doping, which shows that the material is becoming disordered with doping. The broadness or diffusiveness occurs mainly due to compositional fluctuation and structural disordering in the arrangement of cations in one or more crystallographic sites of the structure. This suggests a microscopic heterogeneity in the compound with different local Curie points. The value of γ decreases with an increase in frequency. Figure 13 shows the Curie Weiss fitting of all the samples and calculated γ values for the frequencies 10kHz, 100kHz, and 1MHz.

In order to quantify the rate of change of permittivity with temperature, the derivative of ε has been taken with respect to T and plotted dε/dT with Temperature [Figure 14]. The derivative of ε with temperature indicates that the rate of change of ε at phase transition in the BCTMO1 sample reduced to 1/2, and for the BCTMO2 sample it reduces to 1/8, and for the BCTMO3 sample it reduces to 1/10 of the rate of change of ε for BTO. This reduction of permittivity change at $T_m$ is due to the decrease in the tetragonal percentage, which has a major contribution towards permittivity.

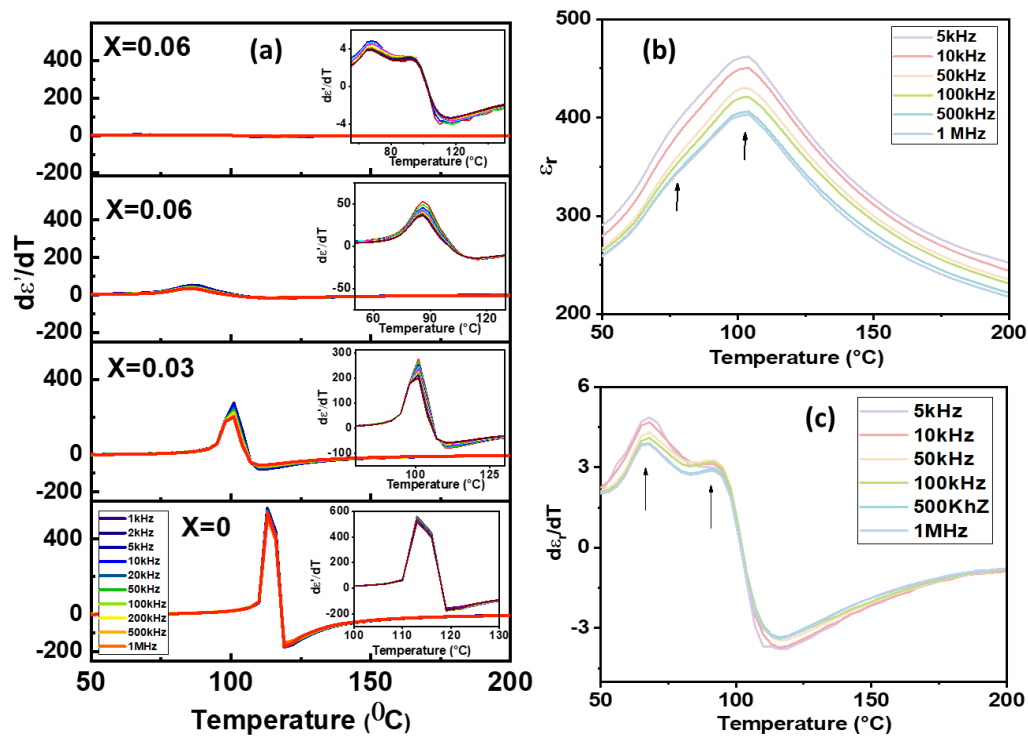

*Figure 14: (a) dε$_r$/dT v/s T plots of all samples and (b)zoomed ε$_r$ (c)zoomed dε$_r$/dT plot of BCTMO3 sample*

Another interesting fact is that the maximum permittivity at the phase transition, ε$_m$ is constant for all frequencies in the BTO sample. However, ε$_m$ value decreases with an increase in frequencies in doped samples [Figure 14]. The change in $\varepsilon_m$ value with frequency is minimal in the case of BTO, whereas this change increases with doping, and the maximum deviation $\varepsilon_m$ with frequency is shown by the BCTMO3 sample. This is due to the increase in space charge polarization and defect dipoles contributions that reduce with frequency, and hence show a reduction in $\varepsilon_m$ as frequency increases. This is also clearer in the derivative plot of $d\varepsilon'/dT$, where the variation of ε shows significant dispersion in the doped samples [Inset figure 14] and is maximum for the BCTMO3 sample. There is an abnormal hump like feature is observed for BCTMO3 before phase transition, as shown in Figure 4.17. For BCTMO1 and BCTMO2 samples, the hexagonal *P6₃/mmc* phase is a minor phase with the major tetragonal *P4mm* phase. However, for BCTMO3, hexagonal *P6₃/mmc* is the major phase, and there is a minor contribution of the tetragonal ferroelectric phase leads to the high diffuseness shown in modified Curie Weiss plots with frequency, a very less $\varepsilon_m$ value (1/5 of the $\varepsilon_m$ of the BTO sample ) as shown in Figure 4.13, a Low value of $d\varepsilon_m/dT$ (~1/10 of the BTO sample), and high dependence of $\varepsilon_m$ with frequency.

When analysing the phase transition temperature of BCTMO3, one can observe a shoulder-like feature just before $T_c$ corresponding to another transition-like feature. This same observation is more visible in the $d\varepsilon_r/dT$ graph. Here, another transition is occurring just before the $T_c$ around ~65-70 °C. The diffused phase transition peaks that occur in these doped BTO are due to the disorder created due to doping, but the secondary peak observed in the BCTMO3 sample before the phase transition is abnormal, and its origin has multiple reasons. Chemical inhomogeneity and strain effects linked to fine-grain size in chemically modified BaTiO$_3$ give rise to diffuse Curie peaks. One of the reasons may be the core-shell structure formed by a strong ferroelectric core surrounded by a weak ferroelectric shell, formed due to the doping. These two structures have different T→C phase transition temperatures. The weak ferroelectric shell T→C transition occurs at a lower temperature than the strong core ferroelectric transition. This kind of shoulder feature also corresponds to the presence of polar nanoregions (PNRs) in the paraelectric phase. This leads to polarisation fluctuations or partial ordering near the transition temperature. It can also be linked to the evolution of ferroelectric domain wall dynamics or intermediate phases, or relaxor-type behaviour induced by dopants. Polar nanoregions (PNRs) or local polarisation clusters exist as precursors to the full ferroelectric-paraelectric transition, reflecting complex local structural and dynamic phenomena preceding the tetragonal to cubic transition.

**Conclusion**

Ba$_{(1-x)}$Ca$_{(x)}$Ti$_{(1-y)}$Mn$_{(y)}$O$_3$ (x=y=0, 0.03, 0.06, 0.09) samples have been investigated for structural polymorphism, band gap tuning, and dielectric anomalies. TEM image of the doped samples indicates the existence of a hexagonal and tetragonal lattice, which are interlinked with each other. XANES analysis confirms the presence of $Ti^{3+}$, $Ti^{4+}$, $Mn^{3+,}$ and $Mn^{4+}$ oxidation states and the structural transformation to a centrosymmetric structure. EXAFS analysis indicates the local structural modification near the Mn and Ti sites with doping. UV-DRS shows a reduction in optical band gap from 3.2 to 2.05 eV and a highest Urbach energy for the x=0.03 sample. The Valence band edge in PES is observed to be shifted towards $E_F$ due to the presence of in-gap states. Density of states calculations reveal a shift of VB and CB with doping in the tetragonal lattice and the formation of defect states with doping in the hexagonal lattice. Electron localization function calculations indicate the lattice contraction with Ca doping in both tetragonal and hexagonal lattices; however, a lattice expansion is observed with Mn doping in both lattices. Dielectric studies show a decrease in the permittivity and an increase in tanδ corresponding to the decrease in tetragonality and increase of the hexagonal phase. A sharp tetragonal to cubic phase transition in BTO transforms to diffuse behaviour with Ca and Mn doping. For x=0.09, the major hexagonal phase sample shows an abnormal shoulder feature before phase transition that may correspond to the ordering of the defect dipoles in the hexagonal phase.

## Acknowledgements

Authors MP and RS would like to thank the Ministry of Education, Government of India, for the Prime Minister Research (PMRF) fellowship. Author DS would like to thank the Department of Science and Technology (DST), Government of India, for the DST INSPIRE fellowship. Author SS would like to acknowledge the DST, Government of India, for providing the funds (DST/TDT/AMT/2017/200). Authors would like to thank Scanning EXAFS Beamline (BL-09) of the INDUS-2 Synchrotron Source, Raja Ramanna Centre for Advanced Technology (RRCAT), Indore, India.